%% file: main.tex
\documentclass[conference]{IEEEtran}
\usepackage[left=1.62cm,right=1.62cm,top=1.9cm, bottom=3.9cm]{geometry}
\IEEEoverridecommandlockouts
\usepackage{amsmath,amssymb,amsfonts}
\usepackage{pifont}

\usepackage[noend]{algorithmic}
\usepackage{algorithm}

\usepackage{graphicx}
\usepackage{textcomp}
\usepackage{lipsum}
\usepackage{subfigure}

\usepackage[backend=bibtex,style=ieee,maxcitenames=6,natbib=true]{biblatex} %added
\usepackage{xcolor}
\def\BibTeX{{\rm B\kern-.05em{\sc i\kern-.025em b}\kern-.08em
    T\kern-.1667em\lower.7ex\hbox{E}\kern-.125emX}}
\begin{document}

%\title{An Opportunistic Hybrid Split-Federated Learning for Mobile UAVs in Heterogeneous Environment}
\title{Opportunistic Transmission of \\ Distributed Learning Models in Mobile UAVs}
%\title{An opportunistic-proactive transmission scheme for Hybrid Learning in Heterogeneous Mobile UAV Networks}
%\title{ Learning in Heterogeneous Environment: an opportunistic hybrid split-federated learning framework}

\author{\parbox{6 in}{\centering Jingxin Li\IEEEauthorrefmark{1}, Xiaolan Liu\IEEEauthorrefmark{2}, Toktam Mahmoodi\IEEEauthorrefmark{1}\\
        \IEEEauthorrefmark{1}King's College London, \IEEEauthorrefmark{2}Loughborough University\\
        E-mail:\IEEEauthorrefmark{1}\{jingxin.1.li, toktam.mahmoodi\}@kcl.ac.uk, \IEEEauthorrefmark{2}xiaolan.liu@lboro.ac.uk}
%\vspace{-1.3\baselineskip}
}

\maketitle
\thispagestyle{plain}
\pagestyle{plain}

\begin{abstract}
In this paper, we propose an opportunistic scheme for the transmission of model updates from Federated Learning (FL) clients to the server, where clients are wireless mobile users. This proposal aims to opportunistically take advantage of the proximity of users to the base station or the general condition of the wireless transmission channel, rather than traditional synchronous transmission. In this scheme, during the training, intermediate model parameters are uploaded to the server, opportunistically and based on the wireless channel condition. Then, the proactively-transmitted model updates are used for the global aggregation if the final local model updates are delayed. We apply this novel model transmission scheme to one of our previous work, which is a hybrid split and federated learning (HSFL) framework for UAVs. Simulation results confirm the superiority of using proactive transmission over the conventional asynchronous aggregation scheme for the staled model by obtaining higher accuracy and more stable training performance. Test accuracy increases by up to 13.47$\%$ with just one round of extra transmission.

\end{abstract}

%\begin{IEEEkeywords}
%UAVs, energy efficiency, split learning, federated learning, opportunistic transmission
%\end{IEEEkeywords}

\input{Section1}
\input{Section2}

\input{Section3}

\input{Section4}
\input{Section5}

\printbibliography %added
\end{document}

%% file: Section1.tex
\section{Introduction}\label{Sec1}
The asynchronous model update in Federated Learning (FL) has received significant attention in recent research. In the traditional FL model, when the server fails to receive timely updates from clients, the aggregation has to be either without consideration of the delayed clients or postponed until responses from all clients are received, both of which degrade the training performance, \textit{e.g.,} reduce the speed of convergence and/or accuracy. Such asynchronicity can occur frequently in scenarios where FL clients are mobile, due to variations in the user connectivity condition. Hence, there is a wealth of literature on techniques to handle delayed or asynchronous model updates. Taking advantage of such literature, in this paper we introduce an asynchronous opportunistic transmission of models from FL clients to the server, which could bring additional benefit in terms of the utilisation of wireless resources.

We focus on a scenario with mobile unmanned aerial vehicles (UAVs) as FL clients. The UAVs, known for their mobility, easy deployment and remote controllability, have been widely used in serving intelligent applications, such as military surveillance \cite{6761569}, metaverse environment creation \cite{MetaUAV}. The UAVs, as aerial users, fly around the target area, collecting data and supporting intelligent applications through wireless networks. Due to the high mobility in 3-dimensional space, UAVs suffer from dynamic wireless transmission, which is easily affected by environmental factors. For example, unexpected moving obstacles can scatter the transmission component, or torrential rain can intercept the transmission. When UAVs act as mobile users, the FL server mandates each UAV user to upload the local model updates within the specified timeframe to conduct the global aggregation, while unstable connectivity often results in failure to do so.

Most existing solutions address the delayed model updates by performing weighted asynchronous aggregation \cite{xie2019asynchronous} \cite{hu2021device} \cite{chen2021fedsa}. Specifically, after receiving the delayed updates, the FL server weights them based on the model staleness and conducts the global aggregation using these weighted updates together with the timely ones. However, as the delayed model updates may greatly vary from the current ones, this weighted-aggregation scheme can still hurt the learning performance, such as accuracy and convergence speed. Hence, rather than passively receiving the staled model updates, proactively transmitting the intermediate local model updates would be a better solution\footnote{We refer to the local model updates sent opportunistically within the local training as the intermediate model updates, to differentiate from the final local model, which is obtained at the end of local training.} and to the best of our knowledge, such approaches are still blank. 

To this end, we propose a novel opportunistic and proactive transmission scheme to tackle the asynchronous model updates issue in FL. During local training, this scheme allows each UAV to upload the intermediate model updates to the server, opportunistically and based on the wireless channel condition. If the final local model updates for a UAV user are delayed or lost, the global aggregation can still be conducted with its corresponding intermediate model updates. We further apply the proposed transmission scheme in one of our previous work, which is a hybrid split and federated learning (HSFL) framework designed for UAVs \cite{liu2022energy}. The HSFL is overall an FL approach while enabling computation splitting and offloading for computing limited UAV users. The two major findings in the evaluation results are as follows,

\begin{itemize}
\item Results show that the proposed transmission scheme outperforms the asynchronous aggregation scheme with a 3.98$\%$ higher accuracy and a more stable training performance. We attribute this improvement to the elimination of the staled model. We also contend that aggregating with intermediate local model updates may be advantageous in the non-i.i.d context since it penalises the local model from overfitting the biased local dataset.

\item The proposed transmission scheme achieves energy efficiency. Specifically, with only one round of intermediate model transmission during the local training stage, the test accuracy on non-iid MNIST data significantly improves up to 13.47$\%$. The transmission of the intermediate model updates is conducted only when the wireless condition is favourable, thus it does not introduce additional communication burdens. 
\end{itemize}
The remainder of this paper is structured as follows. Section II details the scenario and system model. Section III elaborates on the novel model transmission scheme. Section IV presents the simulation results, and finally, section V concludes the paper. 

%% file: Section2.tex
\section{System Model}\label{SystemModel}

\begin{figure}
    \centering
     \includegraphics[width=215pt]{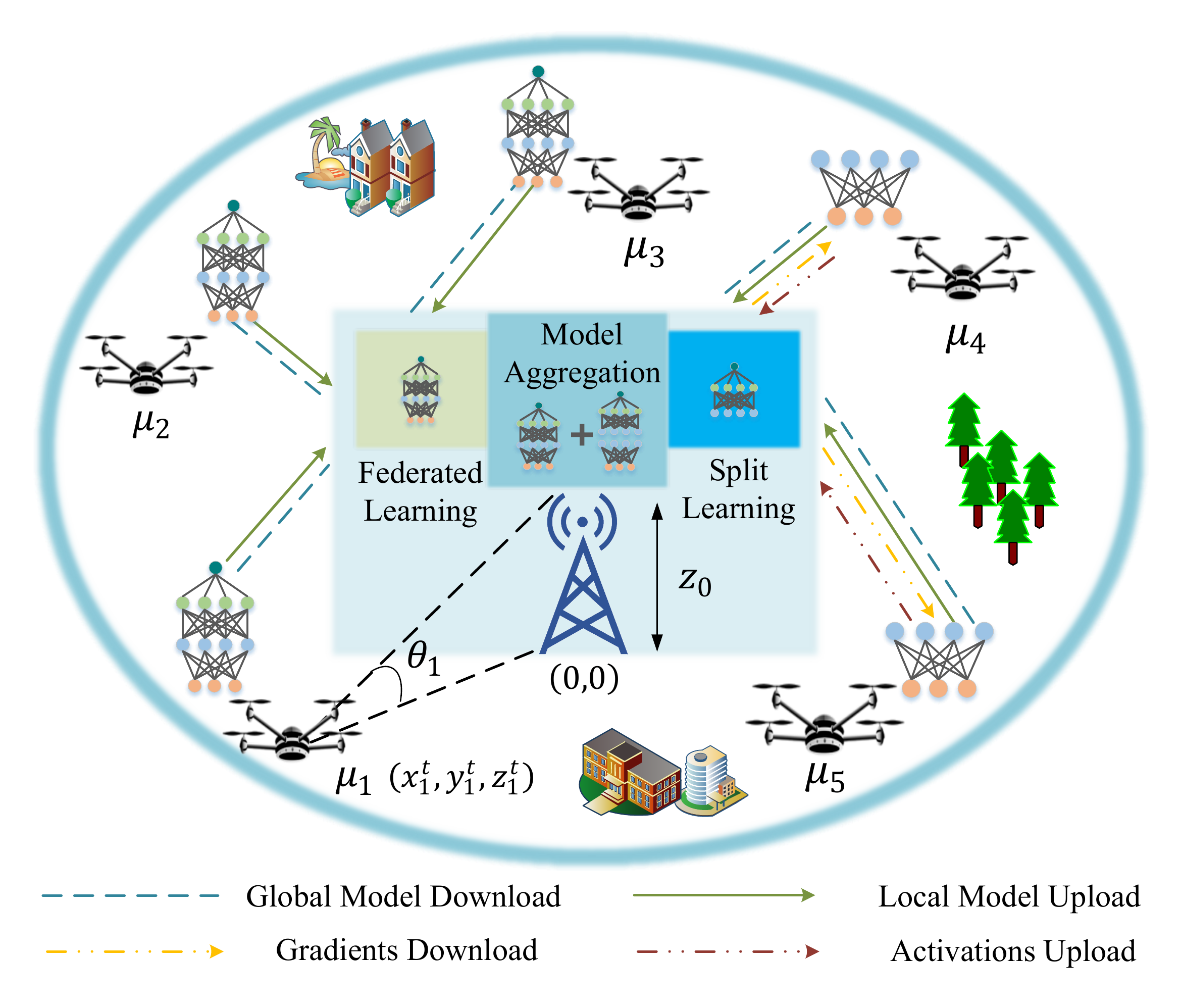}
    \caption{The HSFL framework: Multiple UAVs and a base station in a cell perform model training.}
    \label{fig:UAV_HSFL}
\end{figure}
In this paper, we consider a wireless network, where UAVs fly over a target area to collect data and conduct the assigned learning task. Considering the heterogeneity of the learning environment and energy efficiency of the network, the HSFL framework is applied to train the target DNN, with the collaboration of the UAVs and a base station (BS) server in $B$ communication rounds. Specifically, we have a set of mobile UAVs, $\mathcal{N}= \{\mu_i, i\in N\}$ and the flying route of each UAV \footnote{In the following text, we use the term user and UAV interchangeably.} is assumed unknown to the BS server. Each user owns a local dataset $D_i$, of which the size is denoted as $|D_i|$. 
%There are  $B$ communication rounds in total, and at the beginning of each communication round, a subset of UAVs $\mathcal{K}$ is selected from $\mathcal{N}$, where each selected user is scheduled with either SL or FL based on its characteristics. 

\subsection{Wireless Channel Model}
We consider the wireless channel model as described in \cite{4483593}, \textit{i.e.,}, the Rician fading channel with additional path loss, which combines the expectation of the line-of-sight (LOS) and non-LOS (NLOS) groups. Assuming that the BS locates at the centre of a cell, of which the coordinates are $(0,0,z_0)$, as shown in Fig.~\ref{fig:UAV_HSFL}. $z_0$ denotes the height of BS. The location of UAV $\mu_i$ at time index $t$ is denoted as $(x_{i}^t, y_{i}^t, z_{i}^t)$. Then the distance between UAV $\mu_i$ and BS at time $t$ can be written as,
\begin{equation}
    d_{0i}^t = \sqrt{(x_{i}^t)^2 + (y_{i}^t)^2 + (z_{i}^t-z_0)^2}.
\end{equation}
The elevation angle (in degrees) of $\mu_i$ \textit{w.r.t.} the BS is,
\begin{equation}
    \theta^t_i = \arcsin{\frac{|z_{i}^t-z_0|}{d_{0i}^t}}, \  0^{\circ} \leq\theta^t_i <90^{\circ}.
\end{equation}
Thus, the probability of the LOS link is calculated as in \eqref{eq:Porb},
\begin{equation}\label{eq:Porb}
    P_{LOS\_i}^t = \frac{1}{1 + a_0 e^{-b_0 (\theta^t_i - a_0)}},
\end{equation}
where $a_0$ and $b_0$ are urban environment parameters. Then, the path loss (dBm) can be written as, 
\begin{equation}\label{eq:pathloss}
    PL^t_i = -\frac{\eta_l - \eta_n}{P_{LOS\_i}^t} - 10 \log_{10} \Bigr[\frac{4\pi (d_{0i}^t)^2f}{c}\Bigr]^2 - \eta_n,
\end{equation}
where $\eta_l$ and $\eta_n$ (dBm) are additional path loss for LOS and NLOS link, $c$ is the speed of light and $f$ is the carrier frequency. Then the channel gain between UAV $\mu_i$ and BS at time $t$ is given by,
\begin{equation}
    g_i^t = 10^{\frac{PL^t_i}{10}}\times (v^t_i + s^t_i),
\end{equation}
\begin{equation}
    v^t_i = \sqrt{\frac{K^t_i}{K^t_i + 1 }}, s^t_i = \sqrt{\frac{1}{2(K^t_i + 1)}},
\end{equation}
where $v^t_i$, $s^t_i$ represent the signal amplitude of the LOS component and the scattered path components (\textit{i.e.,} the NLOS links) between the user $\mu_i$ and the BS respectively, with $K^t_i$ (mW) denoting the Rician fading factor. Therefore, the transmission rate of $\mu_i$ at time $t$ can be derived as,
\begin{equation}\label{eq:tran rate}
    r_i^t = n_i B_{uav} \log_2 (1+\frac{g_i^t P_{uav}}{\sigma^2}), \forall i \in N,
\end{equation}
where $n_i\leq 1$ indicates the allocated bandwidth ratio for user $\mu_i$, $B_{uav}$ denotes the total available bandwidth, $P_{uav}$ is the signal power of UAV and $\sigma^2$ is the noise power.

\subsection{Distributed Learning Model}
Here, we consider the learning task of image classification with a DNN model. Specifically, the UAVs and BS server collaboratively train the model to minimise the overall loss function, 
\begin{equation}
    \min_{\omega} \sum_{i\in N} \frac{|D_i|}{D} F_i (\omega),\  F_i (\omega) = \frac{1}{|D_i|} \sum_{(\mathrm{x}_j, \mathrm{y}_j)\in D_i} f_j(\omega),
\end{equation}
where $D$ is the total dataset owned by the users $\mu_i, \forall i \in N$. Additionally, $f_j(w)$ represents the local loss function $\ell(\mathrm{x}_j, \mathrm{y}_j ;\omega)$,
which denotes the loss on sample $(\mathrm{x}_j, \mathrm{y}_j ) \in D_i$ given the model parameter $\omega$. In this work, we use the cross-entropy \cite{de2005tutorial} as the local loss function $\ell(\cdot)$.

\input{HSFL_alg}
Herein, we propose an opportunistic-proactive transmission scheme and apply it to our prior work, the HSFL framework~\cite{liu2022energy}, an energy-efficient learning framework for UAVs. Details of HSFL are summarised in \textbf{Algorithm \ref{alg:HSFL}}. It is worth noticing that HSFL is overall an FL approach yet it enables computation offloading to the edge server to mitigate the computation burden for computing-limited devices, which is referred to as split learning (SL) in HSFL.  To balance energy efficiency and training accuracy, the BS server selects users for training based on their characteristics, including the one-round latency, the diversity of the user resources and energy consumption.

%% file: HSFL_alg.tex
\begin{algorithm}[t]
\caption{Wireless HSFL Framework}\label{alg:HSFL}
\begin{algorithmic}[1]
\STATE \textbf{Initialise:} Global model $\omega$, UE-model $\omega_l$, BS-model $\omega_e$ 
\FOR{$t=1$ to B}
 \STATE Each user $\mu_i$ uploads characteristic info. to BS server
\STATE BS server selects a subset of users $\mathcal{K}$ from $\mathcal{N}$

\STATE BS server schedules each selected user with FL or SL

\FOR{ $i \in K$  in parallel}
      \IF{$\mu_i \in \mathcal{K_F}$ \COMMENT{Scheduled with FL}} 
     \STATE BS server distributes $\omega^{t-1} \to \mu_i$
     \STATE $\mu_i$ computes local model updates with FL
       \ELSIF{$\mu_i \in \mathcal{K_S}$ \COMMENT{Scheduled with SL}}
       \STATE BS server distributes $\omega_l^{t-1} \to \mu_i$
       \STATE BS server initialises $\omega_{ei}^{t-1}$
       
      \STATE $\mu_i$ computes local model updates using SL method
      \STATE jointly with BS server
      \ENDIF
      \ENDFOR
      \STATE BS performs model aggregation with FedAvg  \cite{mcmahan2017communication} $\to$ $\omega^t$
\ENDFOR
\end{algorithmic}
\end{algorithm}

%% file: Section3.tex
\section{Opportunistic-proactive transmission scheme}\label{OpportTrans}
In this section, we propose a novel transmission scheme for the HSFL framework, namely the OPT-HSFL, to mitigate the impact of dynamic wireless conditions on model transmission. As shown in Fig. \ref{fig:opttransmission}, the intermediate model updates are sent from the user to the BS server during local training. When the final model updates are delayed, as for user $\mu_2, \mu_3, \mu_4$, the intermediate model updates received most recently are used for global aggregation. To achieve that, the proposed transmission scheme is divided into two steps: 1) uplink transmission latency relaxation, to warrant each UAV user more transmission budgets; 2) intermediate model transmission during local training. Specifically, we measure the real-time transmission rate, based on the location and the wireless conditions experienced by the UAV. Then we calculate the real-time latency for transmitting the intermediate model parameters. If the induced latency is acceptable regarding the relaxed uplink transmission latency, then the model updates are uploaded to the server. \textbf{Algorithm \ref{alg:OPT-HSFL}} summarises the details of the OPT-HSFL. %At the end of each communication round, if any final model updates are not received by the server in time, the intermediate model updates of the corresponding user would be used for global aggregation. 

\subsection{Uplink Transmission Latency Relaxation}
\input{opt}
The original one-round latency in the HSFL algorithm \cite{liu2022energy} is written as:
\begin{equation}\label{FL_latency}
    \tau_{iF} = \tau_{iF}^{tr} + \tau_{iF}^{ul} \leq \tau_{max}, \mu_i \in \mathcal{K}_F,
\end{equation}
\begin{equation}\label{SL_latency}
    \tau_{iS} = \tau_{iS}^{tr} + \tau_{iS}^{ul} + \tau_{iS}^{dl}\leq \tau_{max}, \mu_i \in \mathcal{K}_S.
\end{equation}
Notice that, $\tau_{max}$ is the maximum one-round latency allowed by the system. Equation \eqref{FL_latency} denotes the one-round latency for users scheduled with FL, $i.e.\  \mu_i \in \mathcal{K}_F$, consisting of the local training time $\tau_{iF}^{tr}$ and the uplink transmission delay, 
\begin{equation}\label{FL_ul}
    \tau_{iF}^{ul} = \frac{m_i^g}{r_{i}^0}.
\end{equation}
$m_i^g$ denotes the size of the local model and $r_i^0$ indicates the transmission rate for user $\mu_i$ at the start of each communication round, which is submitted to the BS server for user selection. Equation \eqref{SL_latency} represents the one-round latency for a user scheduled with SL $\mu_i \in \mathcal{K}_S$, \textit{i.e.,} part of the computation offloaded to the BS server. The latency consists of the local training time $\tau_{iS}^{tr}$, the downlink transmission delay $\tau_{iS}^{dl}$ and the uplink-transmission delay \eqref{SL_ul},
\begin{equation}\label{SL_ul}
    \tau_{iS}^{ul} = \frac{m_i^l +  m_i^a}{r_i^0}.
\end{equation}
$m_i^l$ is the size of the UE-side model and $m_i^a$ is the size of the activations of the cut-layer, which depends on $|D_i|$. Details for calculating $\tau_{iF}^{tr}, \tau_{iS}^{tr}, \tau_{iS}^{dl}$ can be found in \cite{liu2022energy}. Equation \eqref{FL_latency} and \eqref{SL_latency} show that all scheduled users must have one round latency no larger than $\tau_{max}$.
\input{Algorithm}
We introduce a new parameter $\mathrm{b}$ to relax the uplink transmission delay, which concerns only the unilateral transmission from the users to the BS server. $\mathrm{b}$ represents the total number of transmissions sent from the user to the BS server. For example, $\mathrm{b}=1$ means only one model transmission is conducted at the end of the local training while $\mathrm{b}=2$ means that one additional intermediate model transmission is conducted during local training. The updated uplink-transmission latency for FL and SL are now written as,
\begin{equation}\label{Updated_FL_SL_ul}
    \tau_{iF}^{ul} = \frac{\mathrm{b} * m_i^g}{r_i^0},\  \tau_{iS}^{ul} = \frac{\mathrm{b} * m_i^l +  m_i^a}{r_i^0}.
\end{equation}
%\textit{Remark 1:} For FL, the opportunistic transmission can be conducted in parallel with the local training, of which the training time depends on the model size and computing unit capability. Here, we consider the worst latency by adding the transmission time and local computing time together for user selection.

\subsection{Transmission during Local Training}
At the beginning of each communication round, with the updated latency in \eqref{Updated_FL_SL_ul}, the greedy user scheduling algorithm in HSFL \cite{liu2022energy} selects a set of users for training and schedules each user with either FL or SL. For each selected UAV, $i.e.\ \mu_i \in \mathcal{K}$, during the local training, we first calculate the time allowance $\tau_{i\_extra}$ for opportunistic model transmission,
\begin{equation}\label{tau_extra}
    \tau_{i\_extra} = \frac{(\mathrm{b}-1)*m_i}{r_i^0},
\end{equation}
where $m_i$ is the model size, $m_i^g$ for FL and $m_i^l$ for SL. The next step is to decide when to conduct the opportunistic transmission. Herein, we propose to transmit the model updates when $e_t\%(\frac{e}{\mathrm{b}})==0$, where $e_t$ is the local iteration index and $e$ represents the total local epochs. Alternatively, it can be manually set by the system. During these scheduled iterations, the real-time transmission rate $r_i^{e_t}$ is calculated with equation~\eqref{eq:tran rate}. Then the real-time delay for transmitting the intermediate model updates is determined as,
\begin{equation}\label{NewTau}
    \tau_i^{e_t} = \frac{m_i}{r_i^{e_t}}.
\end{equation}
If $\tau_i^{e_t}$ falls within the limit of $\tau_{i\_extra}$, then the current model parameters are sent to the server, after which, $\tau_{i\_extra}$ is updated as follows,
\begin{equation}
    \tau_{i\_extra} = \tau_{i\_extra} - \tau_i^{e_t}.
\end{equation}
These procedures are repeated in the next scheduled $e_t$ until the local training ends. Nevertheless, if $\tau_{i\_extra}$ can not afford the transmission in some scheduled $e_t$, due to the low transmission rate caused by the dynamic wireless condition, then the scheduled transmission is cancelled. 

%\textit{Remark 1:} In the proposed OPT-HSFL, the training performance is improved by trading off extra communication overhead. However, in later sections, we show that the test accuracy can be greatly improved by introducing only a small amount of extra communication overhead.

%% file: opt.tex
\begin{figure}[t]
    \centering
    \includegraphics[width=230pt]{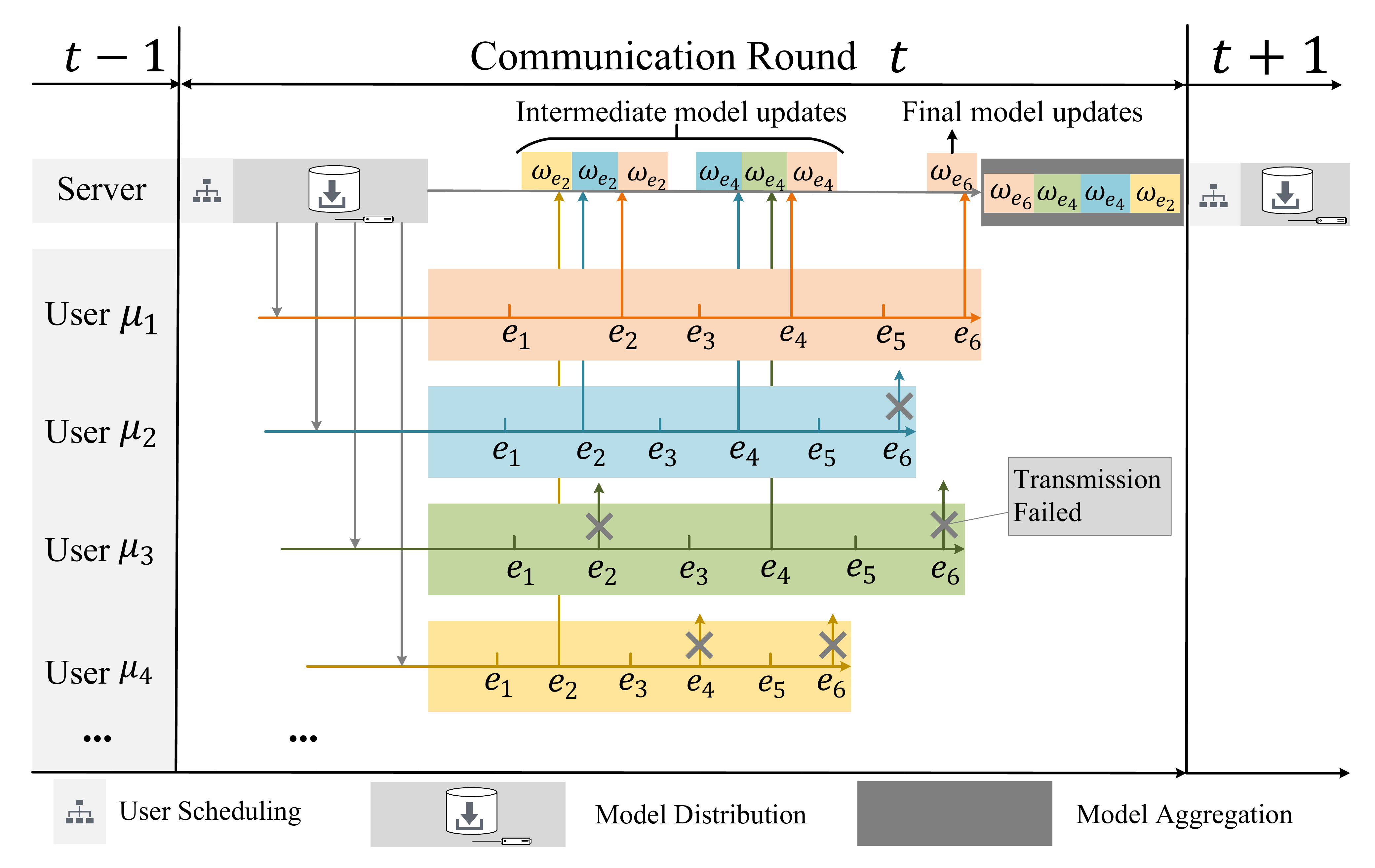}
    \caption{Demonstration of the opportunistic-proactive transmission scheme: $\omega_{e_t}$ indicates the model parameters transmitted at local epoch $e_t$ while the box colour indicates the corresponding user.}
    \label{fig:opttransmission}
\end{figure}

%% file: Algorithm.tex
\begin{algorithm}[t]
\caption{Opportunistic-proactive transmission for HSFL}\label{alg:OPT-HSFL}

\begin{algorithmic}[1]
\STATE \textbf{Initialise:} $\tau_{max}, \omega, e, B, b, \ell r$
\FOR{$t = 1$ to $B$ in parallel, BS Server}
      \STATE $\mathcal{K}=\{\mu_i, i\in K\}\gets$ User selection
      \FOR{ $i \in K$  in parallel}
      \IF{$\mu_i \in \mathcal{K_F}$} \STATE $\omega_i \gets \omega^{t-1}$
       \ELSIF{$\mu_i \in \mathcal{K_S}$}
     \STATE $\omega_i \gets \omega_l^{t-1}$  
    \COMMENT{$\omega_l:$ UE-side model}
      \ENDIF
      \STATE Compute $\tau_{i\_extra}$ \COMMENT{Eq. \eqref{tau_extra}} 
      \FOR{$e_t$ in $e$}
      \STATE$\omega_i \gets \omega_i - \ell r \nabla_{\omega} \ell (\omega_i)$ \COMMENT{Local model updates} 
      \IF{$e_t \% \frac{e}{b}==0$}
      \STATE \textbf{Opportunistic$\_$Transmission($\tau_{i\_extra}, \omega_i$)}
      \ENDIF
      \ENDFOR
      \STATE Upload $\omega_i$ to the BS \COMMENT{Previous $\omega_i$ will be overwritten}
      \ENDFOR
      \STATE Global model aggregation: $\omega^t \gets \frac{1}{|\mathcal{K}|} \sum_{i \in K} \omega_i$
\ENDFOR
\STATE
\STATE\textbf{Opportunistic$\_$Transmission($\tau_{i\_extra}, \omega_i$):}
\STATE Compute $\tau_{i}^{e_t}$ \COMMENT{Eq. \eqref{NewTau} } 
\IF{$\tau_{i}^{e_t} \leq \tau_{i\_extra} $ } 
\STATE Upload $\omega_i$ to the BS \COMMENT{Previous $\omega_i$ will be overwritten}
\STATE Update $\tau_{i\_extra} = \tau_{i\_extra} - \tau_{i}^{e_t}$
\ENDIF
\end{algorithmic}
\end{algorithm}

%\State \textbf{Opportunistic$\_$transmission($\tau_{i\_extra}, \omega_i$):}
%\State Compute $\tau_{i}^{e_t}$ \Comment{Eq. \eqref{NewTau} } 
%\If{$\tau_{i}^{e_t} \leq \tau_{i\_extra} $ } 
%\State  Upload $\omega_i$ to the BS \Comment{Previous $\omega_i$ will be overwritten}
%\State Update $\tau_{i\_extra} = \tau_{i\_extra} - \tau_{i}^{e_t}$
%\EndIf 

%% file: Section4.tex
\section{Simulation Results}\label{Simulation}
\input{SimulationFigure}
In this section, we apply the opportunistic-proactive transmission scheme on HSFL to train a DNN for MNIST image classification \cite{lecun1998gradient} and investigate its training performance in the dynamic wireless environment. We consider a 5-layer convolutional neural model (CNN) consisting of three fully connected layers and two convolutional layers. The training task is conducted with $30$ UAVs and a BS server in $B=100$ rounds. In each communication round, the BS server selects 10 UAVs for training based on their characteristic information. The selected UAVs serve as mobile users, collecting data and conducting local model training. The BS locates at the cell centre, of which the radius is 500 meters, and each UAV randomly flies within the cell during the task training session. The height of the BS is 20 meters, and the vertical flying range of each UAV is 20 to 80 meters. The wireless environment is modelled with Rician fading \cite{913150} with additional path-loss, as described in section \ref{SystemModel}. To impose the wireless dynamics, we update the Rician fading factor $K$ in each local training round by randomly selecting a value from $1.8\sim 5$ dBm. Additionally, the path-loss, as in equation \eqref{eq:pathloss}, also varies in each local epoch. Furthermore, we set the probability of each UAV experiencing a complete communication interruption, caused by sudden weather changes or unexpected moving obstacles, to $30\%$.
\begin{table}[t]
 \caption{Simulation parameters}
 \label{tab:wireless para}
    \centering
    \begin{tabular}{|c|c|}
      \hline
       \textbf{Parameter}  & \textbf{Value}  \\
       \hline
      BS Power $P_{bs}$, UAV Power $P_{uav}$ & 40dBm, 24dBm\\
    \hline 
    Noise Power $\sigma^2$ & -174 dBm \\
      \hline 
      Rician fading factor $K$ & 1.8$\sim$5 dBm\\
      \hline 
      System carrier frequency $f_c$ & 2GHz \\
      \hline 
      BS bandwidth $B_{bs}$, UAV bandwidth $B_{uav}$ & 5MHz, 10MHz \\
      \hline 
      Environment parameters $a_0, b_0$ & 5.0188, 0.3511 \\
      \hline 
      Additional path loss for LOS and NLOS link, $\eta_1, \eta_2$ & 21, 1 dBm\\
     
    \hline 
     Total communication round $B$, Local training epoch $e$    & 100,  6 \\
    \hline 
    Local data batch size, learning rate $\ell r$ & 10, 0.01\\
    \hline 
%    One-round latency limit \tau_{max} & 8\sim 20 s\\
%    \hline 
    \end{tabular}
\end{table}

Table \ref{tab:wireless para} summarises important simulation parameters. Other parameters follow the same setting in \cite{liu2022energy}. Regarding the average communication overhead, we measure the mean of the amount of data transmitted to the server in each communication round, of which the unit is the \textit{megabyte} (MB). We conduct the experiments over three different data distributions to imitate various practical scenarios. The iid and non-iid data distribution follow the settings in \cite{mcmahan2017communication}. Each user only accesses the samples from two classes under the non-iid environment. To set the imbalanced data distribution, we follow the work in \cite{hsu2019measuring} and set $\alpha_d=0.01, \alpha_{imd} = 2$. Smaller $\alpha_d$ indicates higher skewness of sample classes while smaller $\alpha_{imd}$ indicates a more imbalanced size of the dataset.

Fig. \ref{fig:Simulationresults} (a) demonstrates the test loss convergence performances of HSFL with the proposed transmission scheme (solid lines) and without the transmission scheme (dashed lines), where the delayed model updates are discarded. The experiments are conducted over the iid, non-iid and imbalanced distributed data when $\mathrm{b}=2$, \textit{i.e.,} only one extra transmission of the intermediate model updates. As expected, the imbalanced data is most fluctuating, since the heterogeneity exists in both the sample class and the sample size, while the non-iid data is biased only in the sample classes. We can observe that the additional transmission of the intermediate model updates notably reduces the oscillations and converges to a lower loss value for both non-iid and imbalanced data. On the other hand, the iid data is robust to the delayed model updates even without additional transmission. A proper explanation is that since the samples evenly distribute across classes, a few UAVs can provide sufficient knowledge. Therefore, discarding some delayed model updates has little impact on the training performance. Nevertheless, we can still observe from earlier epochs, applying the opportunistic-proactive transmission leads to faster convergence. 

Fig. \ref{fig:Simulationresults} (b) shows the test accuracy of OPT-HSFL under different transmission budgets $\mathrm{b}$. Specifically, we use HSFL with asynchronous aggregation scheme (Async-HSFL) as a benchmark, where the delayed model updates are aggregated with a staleness-based weighting scheme. We set the maximum delay to be 1, which means that the delayed model is received and aggregated by the BS server in the proceeding training round. We follow the polynomial weighting function, $\alpha (t-\tau+1)^{-\mathrm{a}}$, in \cite{xie2019asynchronous}. $t-\tau$ denotes the model delay, which is 1 in this case, and we set $\alpha=0.4, \mathrm{a}=0.5$. Compared with $\mathrm{b}=1$, where the delayed model updates are discarded, the Async-HSFL manages to smooth the oscillations after 50 training rounds yet it suffers from slow convergence. In contrast, even with just one round of intermediate transmission $i.e., \mathrm{b}=2$, the OPT-HSFL converges much faster and smoother while achieving a $3.98\%$ higher accuracy on average than the Async-HSFL, which can be credited to the exclusion of the staled model. From another point of view, when $\mathrm{b}=2$, the global aggregation is conducted with some user updates computed in fewer rounds. In a non-i.i.d. context, each local model tends to fit the biased-local dataset and thus generalises poorly on the balanced test set. By computing fewer rounds of updates, the local model learns less detailed features of the biased dataset and we argue that this leads to better generalisation performance of the global model. 

Fig. \ref{fig:Simulationresults} (c) shows the average communication overhead $\&$ test accuracy of OPT-HSFL on non-iid data under different transmission budgets $\mathrm{b}$. It shows that as the transmission frequency increases, both the test accuracy and the communication overhead increase. From $\mathrm{b}=1$ to $\mathrm{b}=2$, the accuracy is boosted from $86.71\%$ to $94.46\%$ while the communication overhead becomes 2.59 times larger. Although the test accuracy continues to increase as $\mathrm{b}$ further increases, the accuracy improvement is not as significant while the communication overhead becomes much higher. Hence, to balance communication efficiency and accuracy, the labelled point $\mathrm{b}=2$ is an optimal trade-off. Note that as the $\mathrm{b}$ increases, the communication overhead should increase linearly while the orange curve in Fig. \ref{fig:Simulationresults} (c) is not strictly linear. This is due to the randomness in the wireless condition. For example, the Rician fading factor $K$ is randomly selected for each local training round. Thus, in some rounds, if the wireless condition can not warrant the opportunistic transmission within the latency restriction, the transmission would be discarded, resulting in lower communication overhead. The test accuracy slightly decreases from $\mathrm{b}=5$ to $\mathrm{b}=6$ by 0.21$\%$, which can be accounted for by the training randomness as well.

When increasing the one-round latency limit $\tau_{max}$, more UAVs would comply with the latency standard. Thus, the HSFL scheme would allocate more devices to participate in the training. Consequently, both the training accuracy and average communication overhead rise.  Fig. \ref{fig:Simulationresults} (d) shows the test accuracy and average communication overhead over the varying $\tau_{max}$ when $\mathrm{b}=2$. Specifically, when $\tau_{max}$ increases from 8 to 9, the test accuracy increases by 13.47$\%$, but when $\tau_{max}$ further increases, the accuracy improvement is less pronounced. This is because that  $\tau_{max}=8$ is a critical point where only a limited number of UAVs can meet the latency requirement given that the transmission budget $\mathrm{b}=2$.  Nevertheless, as $\tau_{max}$ increases to 9, more UAVs participate in the training, providing sufficient training samples. Therefore, when $\tau_{max}$ further increases to 10 and 11, the improvement becomes comparatively small yet the communication overhead greatly increases. Thus, to achieve energy efficiency, the best trade-off point is indicated by the labelled point when $\tau_{max}=9$.

%% file: SimulationFigure.tex
\begin{figure*}
    \centering
    \subfigure[Test loss]{\includegraphics[width=130pt,height=110pt]{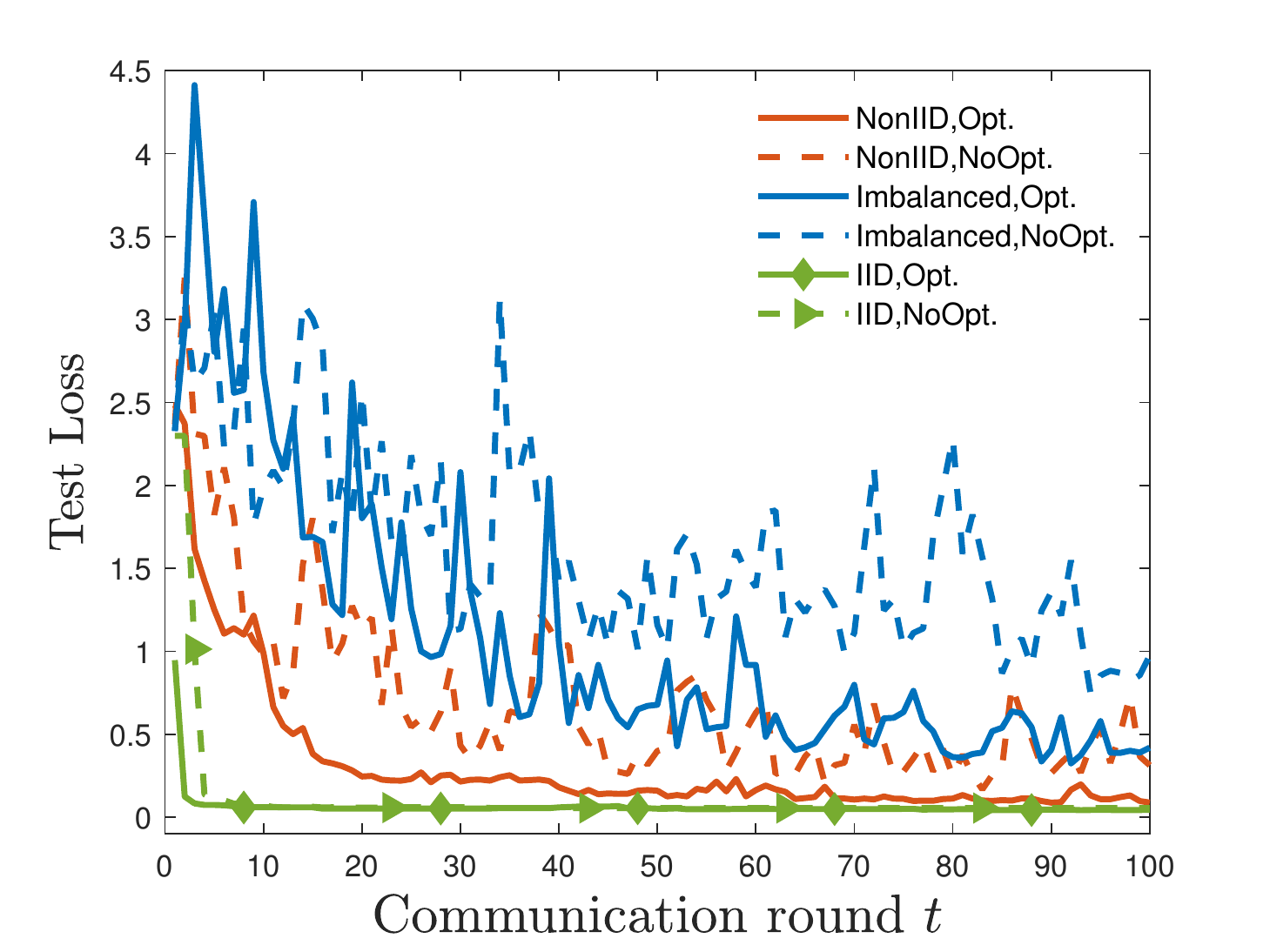}}
    \subfigure[Test Accuracy]
    {\includegraphics[width=125pt,height=110pt]{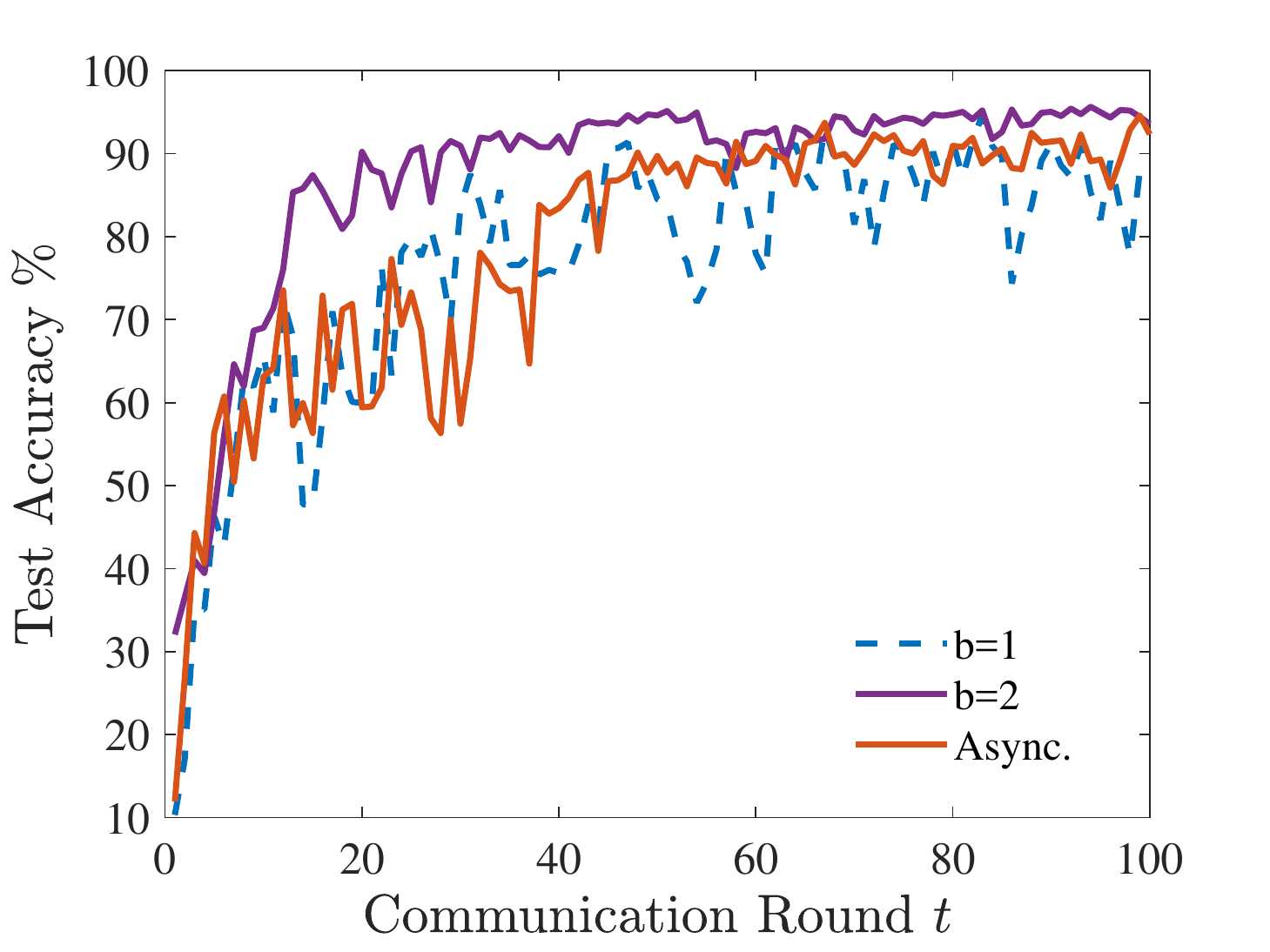}}
    \subfigure[Accuracy$\&$Avg. Overhead]{\includegraphics[width=125pt,height=110pt]{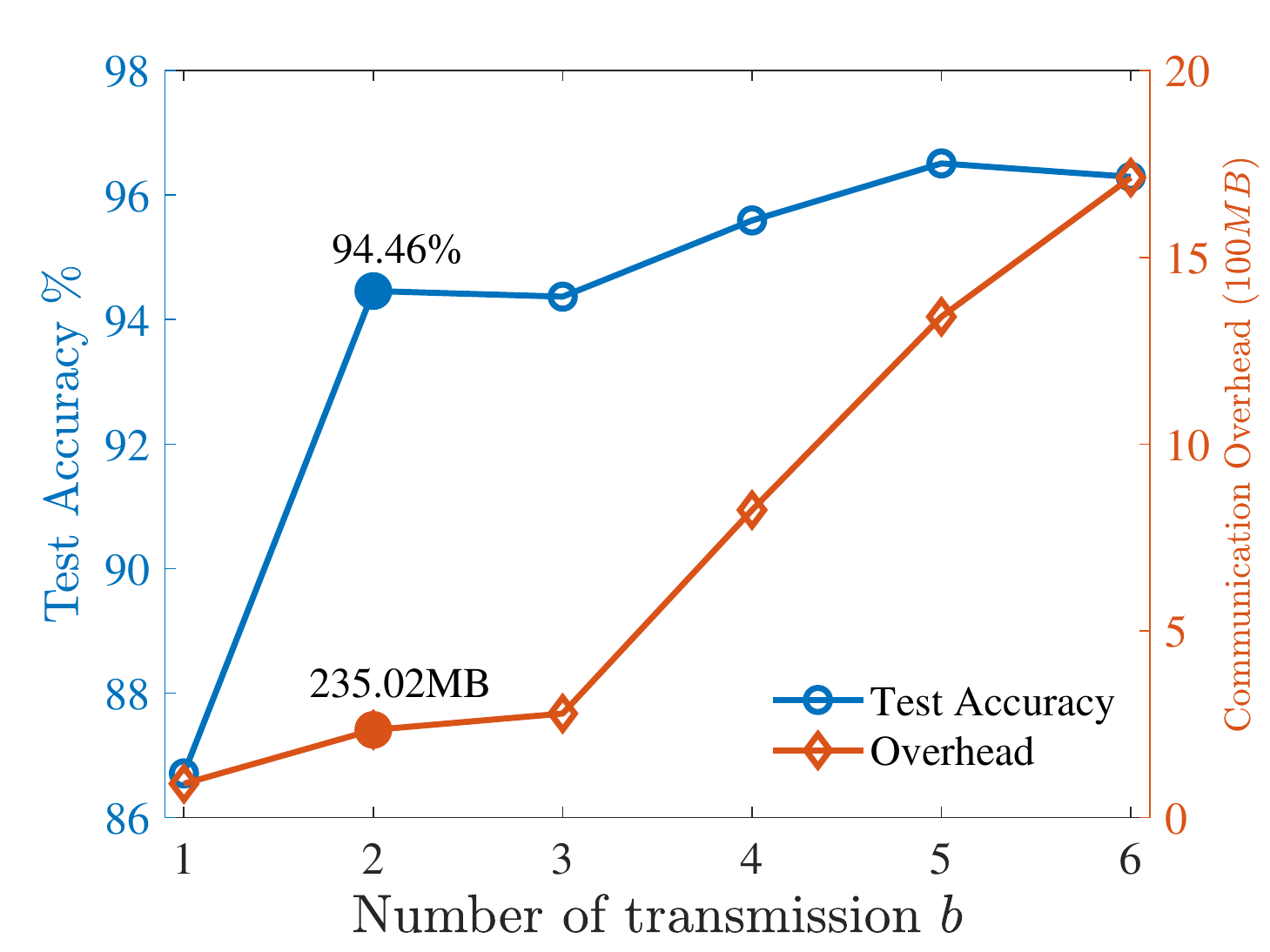}}
    \subfigure[Accuracy$\&$Avg. Overhead]{\includegraphics[width=125pt,height=110pt]{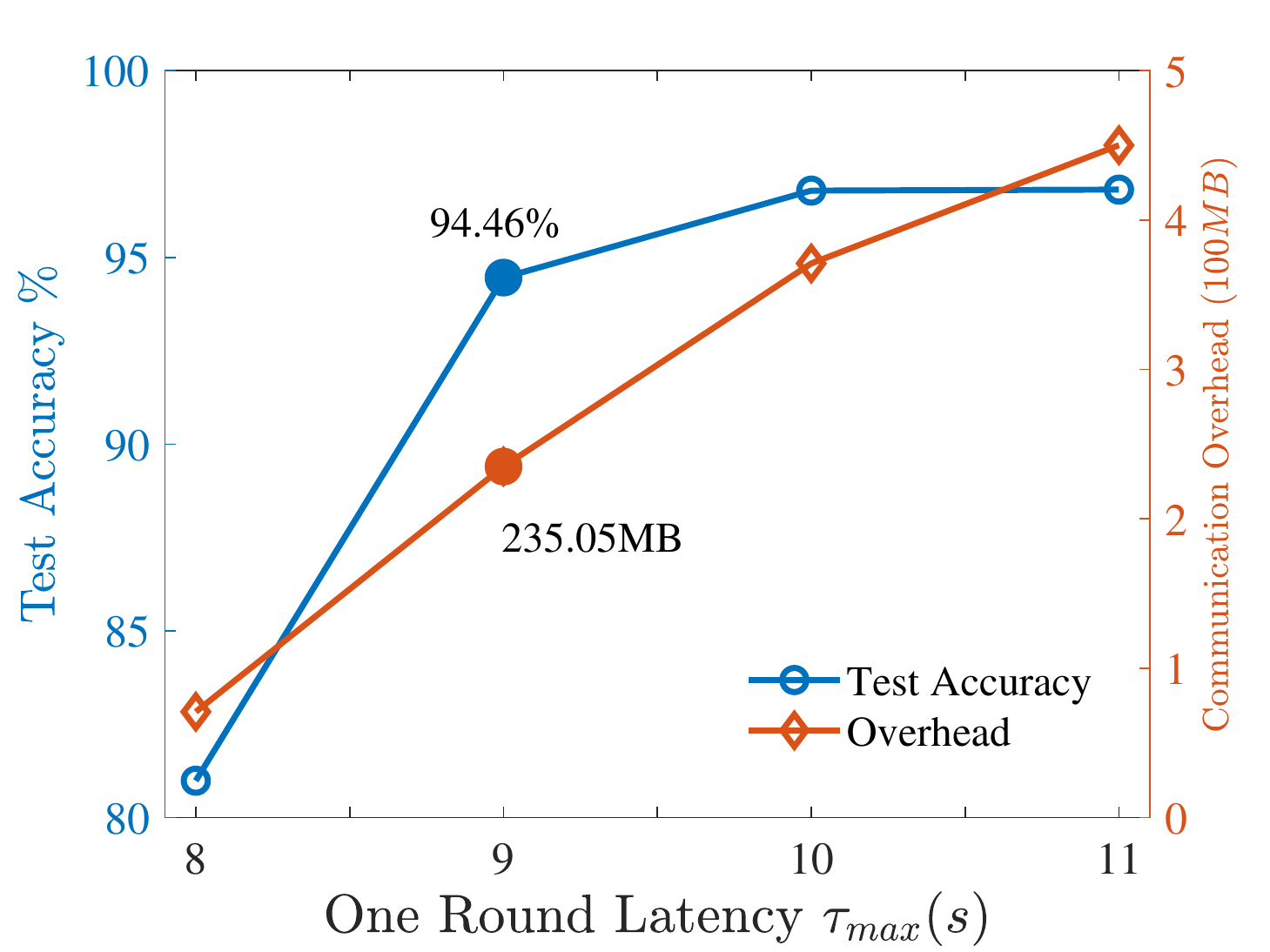}}
    
    \caption{Simulation results: (a) Convergence performance of the test loss for the non-iid, iid and imbalanced data distribution. The solid line represents the OPT-HSFL while the dashed line represents the HSFL scheme without opportunistic transmission, where the delayed model updates are discarded; (b) Convergence performance of the test accuracy for OPT-HSFL and Async-HSFL under the non-iid data distribution; (c) Test accuracy $\&$ Average communication overhead with different $\mathrm{b}$ for the non-iid data; (d) Test accuracy $\&$ Average communication overhead with varied one round latency limit $\tau_{max}$ for $\mathrm{b}=2$ under the non-iid data distribution. Labelled data in (c) and (d): $\mathrm{b}=2$, $\tau_{max}=9s$, accuracy= $94.46\%$, Average communication overhead=$235.02 MB$.}
    \label{fig:Simulationresults}
\end{figure*}

%% file: Section5.tex
\section{Conclusion}\label{Conclusion}
The existing literature on FL asynchronous model updates relies on aggregating the delayed model updates with sophisticated weighting schemes, which could potentially introduce staleness into the global model. In this work, we present a new scheme for handling the such problem in FL. We propose to transmit the intermediate model updates, proactively during local training, and opportunistically depending on the condition of the wireless channel. Simulation results demonstrate the superiority of the proactive transmission, presented in this paper, over the weighted asynchronous aggregation. With just one round of extra transmission, the test accuracy significantly improves by 13.47$\%$, 3.48$\%$, compared with the naive FL and the weighted asynchronous aggregation respectively. Moreover, our proposal obtains a much faster and smoother convergence performance. The advantages of the proposed transmission scheme are more evident in FL applications with longer local training, \textit{i.e.,} large local epochs, since the benefits of intermediate model update can be more significant. An additional observation is that aggregating with intermediate model updates may be advantageous in the non-i.i.d. context since it penalises the local model from overfitting the biased local dataset.

\footnotesize
\section*{Acknowledgements}
This work is contribution by Project REASON, a UK Government funded project under the Future Open Networks Research Challenge (FONRC) sponsored by the Department of Science Innovation and Technology (DSIT).

%% file: main.bib
@incollection{liu2022energy,
  title={Energy Efficient User Scheduling for Hybrid Split and Federated Learning in Wireless UAV Networks},
  author={Liu, Xiaolan and Deng, Yansha and Mahmoodi, Toktam},
  booktitle={Proc. IEEE Int. Conf. Commum.(ICC)},
  year={May 2022}
}

@INPROCEEDINGS{MetaUAV,
  author={Han, Yue and Niyato, Dusit and Leung, Cyril and Miao, Chunyan and Kim, Dong In},
  booktitle={Proc. IEEE Int. Conf. Commun.(ICC)}, 
  title={A Dynamic Resource Allocation Framework for Synchronizing Metaverse with IoT Service and Data}, 
  year={May 2022},
  volume={},
  number={},
  pages={1196-1201}}

@inproceedings{mcmahan2017communication,
  title={Communication-efficient learning of deep networks from decentralized data},
  author={McMahan, Brendan and Moore, Eider and Ramage, Daniel and Hampson, Seth and y Arcas, Blaise Aguera},
  booktitle={Artificial intelligence and statistics},
  pages={1273--1282},
  year={April 2017}
}

@ARTICLE{4483593,  
author={Holis, Jaroslav and Pechac, Pavel},  
journal={IEEE Trans. on Ant. and Prop.},   title={Elevation Dependent Shadowing Model for Mobile Communications via High Altitude Platforms in Built-Up Areas},   
year={Apr. 2008},  
volume={56},  
number={4},  
pages={1078-1084}}

@article{hsu2019measuring,
  title={Measuring the effects of non-identical data distribution for federated visual classification},
  author={Hsu, Tzu-Ming Harry and Qi, Hang and Brown, Matthew},
  journal={arXiv preprint arXiv:1909.06335},
  year={Sep. 2019}
}

@article{lecun1998gradient,
  title={Gradient-based learning applied to document recognition},
  author={LeCun, Yann and Bottou, L{\'e}on and Bengio, Yoshua and Haffner, Patrick},
  journal={Proc. IEEE},
  volume={86},
  number={11},
  pages={2278--2324},
  year={1998}
}

@ARTICLE{913150,  author={Abdi, A. and Tepedelenlioglu, C. and Kaveh, M. and Giannakis, G.},  journal={IEEE Communications Letters},   title={On the estimation of the K parameter for the Rice fading distribution},   year={Mar. 2001},  volume={5},  number={3},  pages={92-94}}

@article{de2005tutorial,
  title={A tutorial on the cross-entropy method},
  author={De Boer, Pieter-Tjerk and Kroese, Dirk P and Mannor, Shie and Rubinstein, Reuven Y},
  journal={Annals of operations research},
  volume={134},
  number={1},
  pages={19--67},
  year={Feb. 2005},
  publisher={Springer}
}

@article{xie2019asynchronous,
  title={Asynchronous federated optimization},
  author={Xie, Cong and Koyejo, Sanmi and Gupta, Indranil},
  journal={arXiv preprint arXiv:1903.03934},
  year={Mar. 2019}
}

@inproceedings{hu2021device,
  title={Device Scheduling and Update Aggregation Policies for Asynchronous Federated Learning},
  author={Hu, Chung-Hsuan and Chen, Zheng and Larsson, Erik G},
  booktitle={IEEE 22nd Int. Workshop on Signal Processing Advances in Wireless Commun. (SPAWC)},
  pages={281--285},
  year={Sep. 2021}
}

@article{chen2021fedsa,
  title={FedSA: A staleness-aware asynchronous Federated Learning algorithm with non-IID data},
  author={Chen, Ming and Mao, Bingcheng and Ma, Tianyi},
  journal={Future Generation Computer Systems},
  volume={July 120},
  pages={1--12},
  year={2021}
}

@INPROCEEDINGS{6761569,
  author={Ma'sum, M. Anwar and Arrofi, M. Kholid and Jati, Grafika and Arifin, Futuhal and Kurniawan, M. Nanda and Mursanto, Petrus and Jatmiko, Wisnu},
  booktitle={2013 ICACSIS}, 
  title={Simulation of intelligent Unmanned Aerial Vehicle (UAV) For military surveillance}, 
  year={2013},
  pages={161-166}}
